# Multiple K Means++ Clustering of Satellite Image Using Hadoop MapReduce and Spark

Tapan Sharma
Research Scholar
ASET, Amity University
Noida, Uttar Pradesh, India

Dr. Vinod Shokeen
ASE, Amity University
Noida, Uttar Pradesh, India

Dr. Sunil Mathur
MAIT, GGS Indraprastha
University, Delhi, India

*Abstract –* **In today's era of big data, with the introduction of high resolution systems, remote sensing imagery is one of the fastest growing fields resulting in a rapid increase in volume of data being generated day by day. To handle such massive volumes of data, a high processing speed has become an indispensable requirement. This is possible with the help of big data platforms such as Hadoop.**

**Clustering of image is one of the important steps of mining satellite images. In our experiment we have simultaneously run multiple K-means algorithms with different initial centroids and values of *k* in the same iteration of MapReduce jobs. For initialization of initial centroids we have implemented Scalable K-Means++ MapReduce (MR) job [1]. We have also run a validation algorithm of Simplified Silhouette Index [2] for multiple clustering outputs, again in the same iteration of MR jobs. This paper explored the behavior of above mentioned clustering algorithms when run on big data platforms like MapReduce and Spark jobs. Spark has been chosen as it is popular for fast processing particularly where iterations are involved.**

## I. INTRODUCTION

Satellite imagery data is growing at a very high rate with the progress in technology. With state-of-the-art technology, petabytes of high resolution data are getting generated on a daily basis. During the past decade or so, this growth has been exponential. It has, thus become important to process and manage such huge amount of data effectively. In the field of remote sensing, deforestation, climate change, ecosystem and land surface temperature are some of the main research areas, where information extraction and the clustering of images play an important role. Since its origin, clustering has been one of the fundamental image processing technique.

Clustering partitions the complete dataset into clusters, or groups by taking into account the similarity among their features. Data points in the same clusters are tend to have similar features while data points possessing different features are grouped in a separate cluster. K-Means, a basic algorithm, is the most popular and widely used clustering technique across all fields, including remote sensing [3]. Although it is the simplest among many of the clustering algorithms, it is the widely used algorithm. Suppose D is a dataset containing n objects. Clustering distributes the objects into k clusters, C1, ...., Ck, where $C_x \subset D$ and $C_x \cap C_y = \phi$ for $(1 \le x, y \le k)$.

Due to the large amount of data generation in satellite imagery, its processing has been a challenge for researchers and industry experts. Hadoop is one of the standard platforms that provides the capability of storing and processing big data in a distributed fashion. It has the capability to distribute the data and process it on distinct nodes in the cluster thereby minimizing network transfers. The Hadoop Distributed File System (HDFS) helps in distributing the portion of the file effectively [4]. In recent times many new frameworks have been built over Hadoop that supports various use cases. Hadoop has now become the de facto framework for distributed programming.

Hadoop has been leveraged by researchers to run the K-means on it [5] [6]. Studies have been conducted to run the algorithm effectively on Hadoop to improve its performance and scalability [1] [7]. Extending the outcomes of these observations, this paper explores the algorithms to run multiple parallel Scalable K-means++ clustering on satellite images for different values of k in the same set of iterations instead of running different iterations for each value of k.

The appropriate k value decides the optimal result of clustering. In real world applications, the number of clusters is usually not known in advance. The one possible way is to run K-means incrementally multiple times to determine the correct number of clusters. The suitable value of k is chosen by evaluating cluster validity index for all the executions. Depending on the clusters found for different values of k, we also validated the appropriate number of clusters required. Major corporate and government agencies such as NASA are already using Spark to process petabytes of data [8]. Therefore, it becomes essential to study and perform such experiments on Spark along with Hadoop MapReduce.

Section 2 of the paper provides the background & related work. Section 3 discusses the Hadoop ecosystem with the basic information about frameworks such as Map Reduce and Apache Spark. Section 4 and Section 5 cover the algorithm that we have used for MapReduce and Spark respectively. Section 6 provides all the details of the experiments and the last section of the paper concludes our findings.

## II. LITERATURE SURVEY

K-Means has been one of the most effective clustering algorithms. It has proven its effectiveness in almost all scientific applications including remote sensing. The groups are created taking similarity of data points into account. The sum of squares of distances (SSD) between points and the corresponding centroid is minimized in order to find the appropriate clusters.

### A. K-Means algorithm

K-Means algorithm has two main phases.
1) *Initialization Phase:* Choose the k data points at random as the initial cluster centroids.
2) *Iterative Phase:*

Step 1: For all points, calculate the distance between each point and the centroids. A data point is allocated to the cluster which has the minimum distance from the point.

Step 2: Find the new centroid point by calculating the mean of all the points in the center.

Step 3: Repeat step 1 and step 2 either for few iterations or till the convergence criteria is met.

The convergence criteria is met when the cluster stabilize. There should be no change in the cluster composition i.e. the points remain in the corresponding cluster for consecutive iterations. The complexity of the algorithm comes out to be $O(nki)$ where $n$ stands for the total points, $k$ is the total partitions (clusters) to be created and $i$ stands for the iterations to be performed [9].

### B. Limitations and Drawbacks

Along with simplicity and effectiveness, the algorithm has certain limitations and drawbacks.

The choice of initial centroids may not result in optimum clustering. If the initial centroids are not chosen wisely then K-Means may converge to just a local optimum.

Secondly, the sequential nature of algorithm makes it challenging to process in parallel. Experiments have been conducted to overcome this sequential nature by running the algorithm in distributed environments.

Thirdly, the value of $k$ needs to be known in advance. K-Means has gone through many improvements with respect to above issues but the core algorithm has remained the same.

### C. Related Work

Scientists and researchers have proposed various algorithms to optimize the initialization phase. Since the initial centroids are chosen at random, one cannot rely on the output of the algorithm for analysis. The greater the number of the initial centroids chosen closer to the actual points, the better is the clustering result.

AlDaoud and Roberts proposed the initialization method based on the density of uniformly partitioned data [10]. The data is partitioned into N cells and the centroids are then chosen from each cell based on the density proportion of that particular cell.

Kauffman also came up with the greedy approach to initialize the K point [11]. In 2007, Arthur and Vassilvitskii proposed the K-means++ clustering algorithm [12]. The paper proposed the careful seeding methodology to select the initial points closer to the optimum result. The initial point is chosen at random. Next subsequent centers are calculated proportional to the closest squared distance from the already chosen center. But this and many other such algorithms suffer from their sequential nature similar to K Means. This iterative nature of the algorithm prevents them from scaling.

Researchers worked on finding parallel implementations of K-Means to find an alternative to sequential nature of the traditional algorithm. With advancement in technology in each and every field, data sets kept on increasing which made clustering algorithms difficult to scale on a single machine even if executed in parallel threads. The horizontal scaling became necessity.

In [5], researchers proposed the K-Means clustering algorithm which ran in parallel based on MapReduce. Zhenhua et al. in [3] ran the MapReduce K-means clustering on satellite images. Taking the help of MapReduce execution framework, the algorithm scaled pretty well on commodity hardware.

Later, [1] came up with scalable K-means++ to optimize it further by sampling more points in each iteration instead of single point. Even though the number of iteration decreases still many iterations are needed. In [6], the paper proposed an efficient k-means approximation with MapReduce. It proposed that a single MapReduce is sufficient for the initialization phase instead of multiple iterations.

In case of MapReduce, multiple iterations of MapReduce jobs are needed to find the appropriate number of clusters. To overcome the iterative MapReduce approach, Garcia and Naldi proposed running the same MapReduce jobs for all the values of *k* [7]. The final output of the MR jobs consists of set of clusters for all the values of *k*. They used the Simplified Silhouette index evaluation to choose the correct value of *k*. They compared the execution results with Apache Mahout for data sets of different sizes generated by the MixSim R package. The paper shows the parallelizing iteration phase for generating clusters. The paper did not conduct the experiment on real data and the initialization phase was not parallelized.

## III. HADOOP ECOSYSTEM

Hadoop is a framework which deals with storage of large amount of data and processing them in distributed manner. A large file is split into multiple parts on different nodes of the cluster. The same task gets executed on each part of the file simultaneously.

Hadoop is based on two core concepts. Hadoop Distributed File System (HDFS) and Distributed Processing Framework MapReduce. After the creation of Hadoop by Doug Cutting, it has evolved a lot and has built a well-integrated ecosystem around its core concepts of MapReduce and HDFS. Hadoop has been used by Yahoo, Facebook and other big companies. While HDFS takes care of the storage of files in a distributed fashion, MapReduce runs the program distributed over the Hadoop cluster.

### A. Hadoop Distributed File System (HDFS)

HDFS works with the help of two types of nodes. Name node and Data node. The Name node is the node in the cluster that stores the details of all the files on the cluster. The Hadoop client talks with this node to read and write from/to data nodes. The name node stores the locations of all the splits of the large file. The data node actually stores the part of the file. The block size is the maximum size of the file portion which can be stored on one node. In earlier versions of Hadoop the default value of block size used to be 64 MB. This has been increased to 128MB now.

### B. MapReduce

On the other hand, a MapReduce layer consists of Job tracker and Task trackers. The job tracker is initialized every time a job starts executing. It is the responsibility of the job tracker to initialize the multiple task trackers on each and every node where the split of the file exists. The task trackers then execute the desired task on the respective nodes. In this way, Hadoop achieves parallel execution on commodity hardware.

The MapReduce paradigm works on the basis of Map and Reduce tasks. A map task takes a <key/value> pair as an input which is decided by the Input Format provided to the MapReduce job. Text Input Format is the default input format which returns a Line Record Reader object [13]. The map task performs some operations and output another form of key-value pairs. Once all the map tasks are completed, the reducer pulls the data from the nodes where the map tasks are running. The reducer then takes the <key, value-list> as an input wherein the values corresponding to a single key form the value-list. All the values of the key go to the same reducer. The reducer can then perform the required operations and write to the output in key-value pairs which is by default a text format.

For reading binary data there are other file formats specific to Hadoop such as Sequence File Format which stores data in binary key-value pair. Since our dataset deals with small and medium sized images, Sequence Files are appropriate for our experiment [4].

### C. Apache Spark

Similar to Hadoop MapReduce, there is another clustering computing framework, called Spark [14]. Spark can be deployed on HDFS as well as standalone. Spark has some in-memory processing capabilities, although it doesn't store all the data in the memory. In our experiments using Spark, we have deployed it over HDFS.

The core concept of Apache Spark which has helped Spark in achieving high performance is Resilient Distributed Datasets (RDD). RDDs are distributed, lazy and can be persisted. Since the data is distributed on HDFS over multiple nodes the processing can happen on individual nodes minimizing the transfer of data over the network. This is similar to the MapReduce approach. Another property is laziness. Spark reads all the operations to be executed for the particular output and doesn't execute them until the output is explicitly asked. The operations are not executed individually which means the data need not to be stored on disk in order to pass it to the next operation. RDD can be persisted in the memory as well on the disk. Keeping the RDD in the memory makes the process faster as the same RDD can be used multiple times without computing it time and again.

RDDs are meant to be designed for iterative algorithms and batch processing which applies the same operation on all the records of the data set. In our case, the implementation of the various forms of the K-Means algorithm is iterative in nature. Such algorithms can be easily benefitted from the RDD approach. Jeremy Freeman has put focus on Apache Spark while discussing the open source tools for large scale neuroscience [15]. The paper has dealt with imaging in neuroscience and states that the performance of image processing can be increased multi-fold by using Spark.

## IV. METHODOLOGY

For this paper, we have implemented the MapReduce to find the optimum value of $k$ by executing the K-means algorithm with modified values of K in the same set of MapReduce iterations. Figure 1 depicts the schematic diagram of the methodology that we have followed.

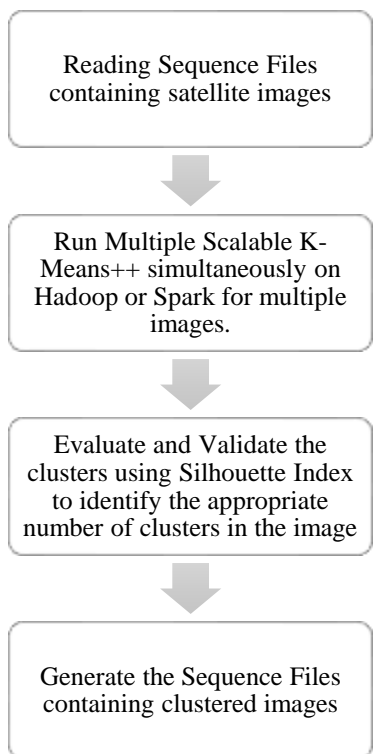

Figure 1. Methodology

We have divided our experiments into three phases viz. *Initialization* phase, *Clustering* phase and *Validation* phase. All the three phases run same MR iterations for different $k$ values. This helps in drastically reducing the time spent in running different iterations for different values of $k$.

For the first phase, we have applied the scalable k-means++ to initialize the clusters for different $k$ values. Although [6] has better performance than [1] but since Spark has in built support for the scalable K-means++, therefore for MapReduce we implemented scalable K-means++ initialization.

The second phase is simple K-means clustering which helps in returning the clusters. This means for values of $k$ equal to two the iteration would return two clusters. For value of $k$ equal to four it would return four clusters, and so on.

The third phase validates and suggests the appropriate value of $k$ which best suits the selected satellite image. For this phase, the Simplified Silhouette-index method has been used [2].

### A. Multiple K-means MapReduce (MKMeansMR)

As discussed in earlier sections, Sequence Files are created for processing of satellite images. A bird's eye view of the steps that we have taken for our experiment is shown in Figure 2.

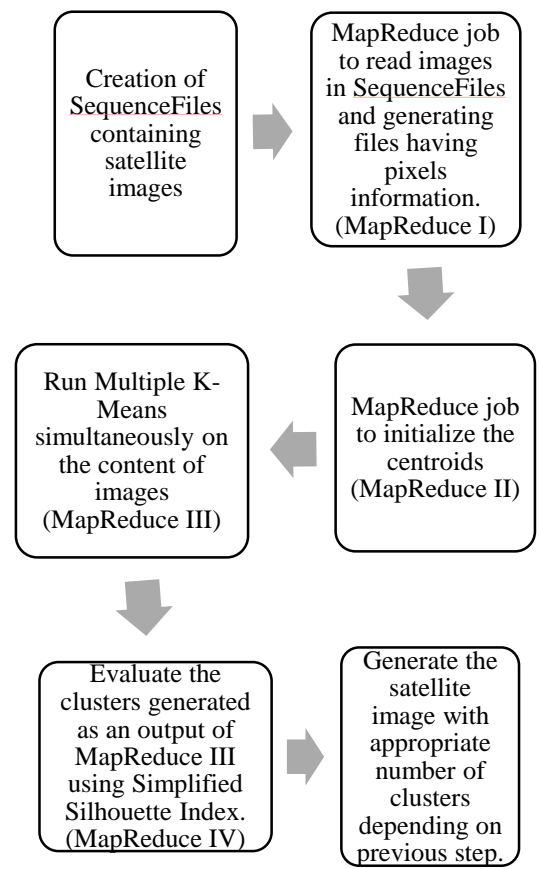

Figure 2. MapReduce Experiment

Taking the benefit of MapReduce, the Scalable K-Means++ algorithm is executed simultaneously for multiple values of k instead of running MR jobs separately for each value of *k*. In our experiment, we have also executed validation algorithm Simplified Silhouette Index for multiple values of *k* so that to identify the appropriate value of the number of clusters. This section covers the description of the three algorithms, all of which work for multiple values of *k*.

*1) Reading RS Image*

The first MapReduce program (MR I) takes Sequence File as an input with the file name as key and Bytes Writable as value. The Bytes Writable object is converted to Buffered Image. The pixels of the image are then traversed and the output of this mapper-only job contains the x and y coordinate of the pixel along with its RGB value. Each pixel in a remote sensing image is represented by its RGB value.

*Mapper:*

*a) Pre-requisites:*

Sequence File [key=filename (Text), value= Image (Bytes Writable)]

*b) map():*

- Convert Bytes Writable to Buffered Image
- For all pixels, write the x and y coordinates of the pixel to the context along with its color

*2) Scalable K-Means++ initialization MapReduce*

In the second MapReduce (MapReduce 2 or MR2), we implemented the Scalable K-Means++ initialization method [1]. We randomly chose a point in the image as the first centroid.

A file, *initialpoints.txt* contains all the centroids' locations produced during this phase. In the setup of MapReduce 2, we read this file and populate the centroid list for our mapper. The file created after MapReduce 1 is sent as an input to MapReduce 2 which contains all the pixels' information.

This MapReduce job is run iteratively for only a few rounds as they should be enough for good results [1]. For each run of MR2, new centroid locations are calculated and the *initialpoints.txt* file is updated accordingly. The MapReduce 2 continues to execute a few rounds to get the *k* initial centroids from C sampled clusters.

When it comes to K-means and finding the similarity, generally CIELAB is used rather than RGB as It is designed to approximate human eye [3] [16]. CIE-(L*a*b*) color space is the form of CIE XYZ. Here 'L*' provides information of a luminosity, 'a*' represents the chromaticity layer which indicates where the color lies on the red-green axis. On the other hand 'b*' tells about the color falling along the blue-yellow axis [17].

Each pixel is then represented by a* and b* values. Euclidean distance formula is then used to calculate the difference and thus similarity between pixel colors.

*Mapper:*

*a) Pre-requisite:*

initpoints.txt: centroids file [Different files for different value of k]

*b) Input:*

File containing all pixels' information(x, y, color)

*c) setup:*

Read initpoints.txt and populate list of cluster centroids

*d) map():*

- For each pixel:
- Loop through all the cluster centroids and calculate the Euclidean distance between the Lab value of centroid and the input pixel. The lowest distance is considered to the nearest distance. An object of Initial Points Detail is created which has pixel coordinates, its color and square of nearest distance, $d^2(x, C)$, where x is the point and C is the set of points in the cluster.
- A key containing *k* value or partition Id and the Initial Point Details object as a value is written in a context.
- End of for

*Reducer:*

Multiple reducers are invoked depending on the number of partitions, *k*.

*a) Input:*

List of Initial Point Details object.

*b) reduce():*

- For all the initial point details object:
- Calculate the sum of minimum distances $\varphi_x(C)$
- End of for
- Clusters are drawn depending on the oversampling factor *l* and using the formula, $l*d^2(x,C) / \varphi_X(C)$ as per scalable K-means++ [1]

*c) cleanup():*

The content of initialpoints.txt is updated to include the new centroid point found in step 5. Since a file cannot be updated in HDFS, a new file is created keeping the previous content as it is.

During the last iteration, *k* clusters are sampled from the list of *C* sampled clusters. We can get one centroid file corresponding to each value of *k*.

*3) Multiple K-means clustering*

The output of MR2 generates multiple set of files containing all the *k* initial centroids. Since we are running multiple k-means, therefore different copies of the file of initial centroids is created depending on the number of clusters to be formed. For an example, for calculation of 5 clusters, a file should contain 5 initial centroids. Similarly for 7 clusters, the details of 7 initial points should be present.

*Mapper:*

  *a) Pre-requisite:*

Centroid files (*centroid_k.txt*) for each value of k. Each line of these files contains a clusterId, pixel coordinates and color. This is the output of MR 2.

  *b) setup():*

Read all centroid files and store information in HashMap (partitionCentersHashMap) where key is the value of k and value contains the list of centroids against k.

  *c) map():*

This method is called for every pixel coordinate.
- For each key in partitionCentersHashMap:
- Loop through all the clusters.
- Euclidean distance is calculated between the pixel color and the cluster color.
- Find nearest cluster, create a *PixelWritable* object containing pixel information along with the cluster in which it falls during this job run.
- *Context Output*: k as the key and *PixelWritable* object.

*Reducer:*

Number of reducers are set equal to the value *k*. This receives all the pixels which have same *k* value.
- Pixel information is iterated and new values of Clusters are found.
- Update the centroids files (*centroid_k.txt*) as per this updated information

4) *Validating clusters*

The Silhouette Index algorithm is used to validate the consistency of data clusters. The purpose of this algorithm is to tell the cohesion among the objects in the cluster. More the cohesion, better the clustering. In our case, we needed to have the algorithm that could be fit within the restrictions of MapReduce. For this, [7] chose Simplified Silhouette Index (SSI) which is one of the variants of the original Silhouette Index. The original Silhouette Index approach calculates all the distances among all the data points. In SSI, the approach is simplified and the distance is calculated between the centroid and the data point.

MapReduce 4 deals with validating the clusters for each partition.

*Mapper:*

  *a) Pre-requisite:*
- File containing all the partitions and centroids corresponding to those partition. Each line contains *partition Id*, *cluster Id*, cluster's X coordinate, cluster's Y coordinate and cluster color.
- File containing the list of points along with the cluster and partition information. Each line contains the *partition Id*, *cluster Id*, *cluster Color*, *point X*, *point Y*, *point Color*.

  *b) setup():*

Loop through the first file and create a hashmap (partition Center Map) containing partition Id as key and cluster information as value.

  *c) map():*

This is called for each line of the second file.
- Each line of the file corresponds to the *partitionId* (k-value) and the pixel information.
- Depending on the pixel and its designated cluster, calculate the distance between the two (*distance*)
- Also, from all the clusters in the corresponding partition find the cluster with minimum distance (*minClusterDistance*) from the point.
- Apply simplified silhouette index formula- (*minClusterDistance* – *distance*) / *minClusterDistance*.
- Emit the SSI with *partitionId* as the key.

*Reducer:*

One reducer for each partition is instantiated.
- Calculate sum of all SSI and divide it by the number of pixels.
- Output the partitionId along with corresponding SSI.

*B. Spark Multiple Scalable K-means++*

Apache Spark is the open source framework which fits over the HDFS layer in the Hadoop ecosystem. It is easy to do development in three languages viz. Scala, Python and Java. It can be seamlessly integrated with different file systems such as HDFS, local file system and Amazon EC2 [18].

*1) Process RS image Spark job*

Spark comes with different libraries meant for special purposes. These include Spark SQL for structured data processing, GraphX for graph processing and Spark Streaming for streaming. Similarly, Mllib provides out-of-the-box capability for machine learning algorithms.

Mllib provides support for various clustering algorithms out of which K Means is one of them [14]. The Mllib library provides implementation of K means || [1]. In our experiment we have used Mllib classes in order to run Scalable K-means++ on Spark.

The pseudo code of the Spark job is described below.

  *a) Read sequence file:*

Read Sequence File using spark context and store in Java Pair RDD.

  *b) Apply transformation:*

Apply transformation to return the list of *PixelInformation* object for each file. PixelInformation class contains the pixel coordinate and its color.

  c) *For each file:*
- Convert RGB to Lab value and create a vector of a* and b* values.
- *Parallelize* the list of Lab value vectors to create JavaRDD.
- For each value of *k*:
    o Run *KMeans.train* on JavaRDD to return *model*
    o Store model's cluster centers in objects of *Cluster Info.*
    o Parallelize the list of *Cluster Info* to create an RDD and store it to a file. The columns stored in the file would be *partition Id*, *cluster Id*, *cluster Color*.
    o For each color pixel, predict the cluster index from the model.
    o Create a *Point Information* object using this information
    o Save the information gathered into the text file where each line would have these columns: *partition Id*, *cluster Id*, *cluster Color*, *point X*, *point Y*, *point Color*.
- End for each value of k.

Our Spark program had three steps similar to MapReduce. As the pre-requisite, satellite images were converted into a sequence file. In a sequence file, these satellite images were stored as values in key-value pairs with the file name as the key. Sequence Files were then read as byte arrays and converted into Buffered Image.

All the images in the sequence file were then read and the information of each pixel was stored in the class Pixel Information containing the pixel coordinates and its color. The list of pixel information for each file is then stored in Java Pair RDD. The algorithm described above is executed. As mentioned in the algorithm, K-Means implementation of MLLib was applied iteratively for every value of *k*. We could not run multiple Scalable K Means++ on our images in parallel as nested RDDs cannot be used inside the transformations [19]. The algorithm generated two text files. First file contains the list of clusters with its information such as *clusterId* and the color of the cluster. The second file contains the information of each and every pixel in the image along with the cluster and the partition it belongs to.

2) *Validating clusters*

The text files generated in the first algorithm acted as an input for the second algorithm. We implemented the Simplified Silhouette Index on Spark using Java. The SSI is executed for every value of *k* in parallel. With the help of the Silhouette Index the appropriate number of clusters for the satellite image can be computed. If the user wants to know the number of clusters in which the satellite image needs to be partitioned, then Simplified Silhouette Index is applied on the points of the clusters. This needs to be run on each value of k. Clustering is considered better when the value of SI is small. Instead of iteratively computing the value of Silhouette Index for different values of "k", our program execute SSI in parallel for multiple values of k.

The output files of first algorithm are the pre-requisites for this job. The first file contains the cluster information. The second file contains the pixels' information for corresponding cluster.

  a) *Setup Phase*
- Read the file containing the partitions along with the cluster information produced as an output of the first algorithm.
- Another map transformation is applied over the RDD of the previous step which populates the HashMap containing *partition Id* and its list of clusters.

  b) *Computing Phase*
- In the first step, read the content of the second file generated as an output of Algorithm 4. Create JavaRDD of Point Information.
- For each partition:
    o Get list of clusters. Apply map transformation on RDD of step 1 to calculate the silhouette values of every pixel.
    o Calculate the sum of silhouette values using reduce transformation on RDD of previous step and then compute the Silhouette Index of the particular partition.

V. EXPERIMENTS

The experiments were conducted on 5 node cluster with one master and 4 data nodes. We created the cluster of Apache Hadoop 2.6 with a replication factor of 3, block size of 128.

The same cluster was used for both MapReduce and Spark programs. We configured Spark 1.5 for our experiment. For MapReduce we configured 8GB memory for map and reduce. Similarly for Spark, executor memory was 8GB with 4 executor workers running per data node.

Our data set included satellite images varying sizes from 481KB to 14.1MB with number of pixels in the range of 2 million to 14 million. Multiple Sequence Files were created with the dataset varying from 1GB to 4GB. We ran our experiments for three values of *k* i.e. for 5, 6 & 7 clusters. The algorithm can be applied for any number of files and for any values of *k*.

### A. Clustering Phase

We measured statistics for speedup and scale-up. For speedup we have fixed the input data set and increased the number of nodes in the clusters varying from 2 nodes to 4 nodes. Similarly for scale up, we increased both input dataset and the cluster size. Scale-up helps in understanding the behavior of the algorithm if both data and cluster size increase.

Figure 3 shows the graph of speedup values for MaReduce jobs indicating that the algorithm speedup performance was good on Hadoop MapReduce.

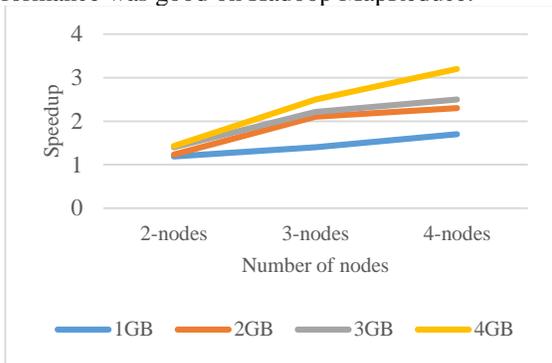

Figure 3. MapReduce Clustering Phase Speedup

For scaleup, we doubled the input data set size and also doubled the number of nodes in the cluster. The input datasets were 1GB, 2GB and 4GB running on 1, 2 and 4 data nodes. The values came out to be 1.2 and 0.9.

Figure 4 shows the graph for Spark jobs running for input datasets of varying sizes. The Spark job speedup had a decent speedup performance.

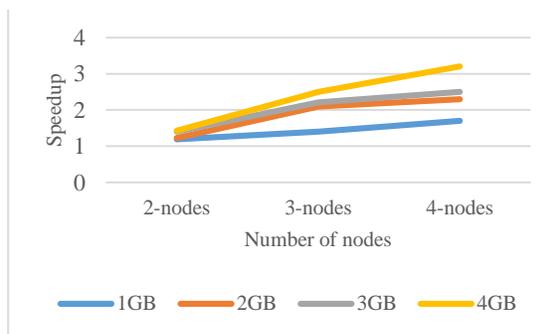

Figure 4. Spark Clustering Phase Speedup

Again, the scaleup was calculated with the same dataset and cluster size combination as we did for MapReduce. The values came out to be 1.16 and 0.92.

### B. Multiple Validation Phase

Similar to clustering phase, we measured the performance of Silhoutte-Index Validation Algorithm for MapReduce and Spark. The behavior of validation phase was similar to clustering phase. The speedup performance was nice for MapReduce. Figure 5 shows the relative graph for the speedup values for different data sets.

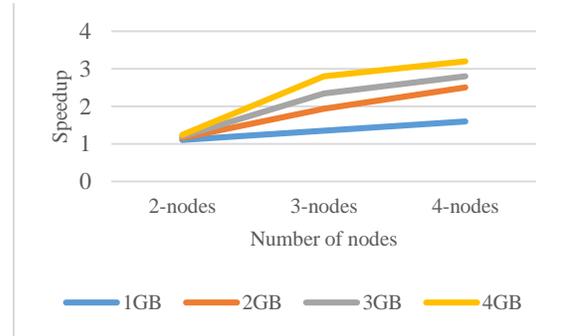

Figure 5. MapReduce Validation Phase Speedup

The scale up values for MapReduce job was measured at 1.12 and 0.74.

Figure 6 shows the speedup performance of Validation phase running on Spark. The validation phase speedup performance looked better than clustering phase. The scale up values were measured at 1.15 and 0.81.

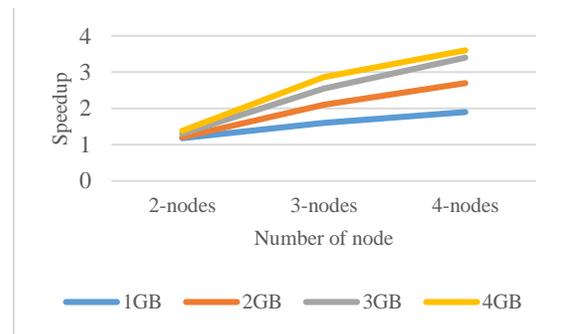

Figure 6. Spark Validation Phase Speedup

CONCLUSION & FUTURE SCOPE

In our experiment we performed K-means clustering simultaneously for multiple values of K with Scalable K-means++ initializations. After that we also computed Simplified Silhouette Index in parallel for multiple partitions with the help of MapReduce and Spark.

We measured speedup and scaleup values for different combinations of data sets, the number of nodes in the cluster. For both clustering and validation phases, the algorithms' speedup performance was decent. The experiments showed that the algorithm can scale well on MapReduce as well as on Spark.

Our cluster environment was small due to which we could not process larger files. In our next experiments, we are planning to use larger image files ranging up to 1GB and a bigger cluster of 16-32 nodes so that a single file will be distributed over multiple nodes. We may have to modify the algorithms in order to exploit MapReduce and Apache Spark's distributed processing capabilities in a better way.

## REFERENCES

[1]  B. Bahmani, B. Moseley, A. Vattani, R. Kumar and S. Vassilvitskii, "Scalable k-means++," in *International Conference on Very Large Databases*, 2012.

[2]  L. Vendramin, R. . J. G. B. Campello and E. R. Hruschka, "Relative Clustering Validity Criteria: A Comparative Overview," *SIAM,* 2010.

[3]  Z. Lv, Y. Hu, Z. Haidong, J. Wu, B. Li and H. Zhao, "Parallel K-Means Clustering of Remote Sensing Images Based on MapReduce," in *2010 International Conference on Web Information Systems and Mining, WISM 2010*, 2010.

[4]  T. White, "Hadoop I/O : File Based Data Structures," in *Hadoop - The Definitive Guide*, O'Reilly.

[5]  W. Zhao, H. Ma and Q. He, "Parallel K-Means Clustering Based on MapReduce," in *CloudCom*, Beijing, 2009.

[6]  Y. Xu, W. Qu, G. Min, K. Li and Z. Liu, "Efficient k-Means++ Approximation with MapReduce," *IEEE TRANSACTIONS ON PARALLEL AND DISTRIBUTED SYSTEMS,* vol. 25, no. 12, December, 2014.

[7]  K. D. Garcia and M. C. Naldi, "Multiple Parallel MapReduce k-means Clustering with Validation and Selection," in *Brazilian Conference on Intelligent Systems, IEEE*, 2014.

[8]  C. Mattmann, "SciSpark: Interactive and Highly Scalable Climate Model Analytics," 2015. [Online]. Available: http://esto.nasa.gov/forum/estf2015/presentations/Mattmann_S1P8_ESTF2015.pdf.

[9]  J. Han, M. Kamber and J. Pei, "Partitioning Methods," in *Data Mining Concepts and Techniques*, Elsevier.

[10]  M. AlDaoud and S. A. Roberts, "New methods for the initialization of clusters," *Pattern Recognition Letters,* pp. 451-455, 1996.

[11]  L. Kaufman and P. J. Rousseeuw, Finding Groups in Data: An Introduction to Cluster Analysis, Wiley, 1990.

[12]  D. Arthur and S. Vassilvitskii, "k-means++: The advantages of careful seeding," in *SODA, Proceedings of the eighteenth annual ACM-SIAM symposium on Discrete algorithms*, 2007.

[13]  D. Miner and A. Shook, MapReduce Design Patterns, O'Reilly.

[14]  "Apache Spark," [Online]. Available: http://spark.apache.org/docs/1.3.0/mllib-clustering.html#k-means.

[15]  J. Freeman, "Open source tools for large-scale neuroscience," *Current Opinion in Neurobiology,* vol. 32, p. 156–163, 2015.

[16]  "Wikipedia," [Online]. Available: https://en.wikipedia.org/wiki/Lab_color_space.

[17]  "Color Management Information from Phil Cruse," [Online]. Available: http://www.colourphil.co.uk/lab_lch_colour_space.shtml.

[18]  A. Spark, "Apache Spark," [Online]. Available: http://spark.apache.org/docs/1.3.0/index.html.

[19]  "Apache Spark JIRA," [Online]. Available: https://issues.apache.org/jira/browse/SPARK-5063.